\documentclass[aps,prl,twocolumn,superscriptaddress,amsmath,amssymb]{revtex4-1}

\usepackage{graphicx}
\usepackage{subfigure}
\usepackage{multirow}
\usepackage{natbib}
\usepackage{array}
\usepackage{multirow,amssymb,amsbsy,amsmath,epstopdf}
\usepackage{color}
\usepackage{ulem}

\usepackage[colorlinks,citecolor=blue]{hyperref}

\begin{document}
%\title{Recycling Bell nonlocality with projective measurement and shared randomness}
\title{Experimental recycling of Bell nonlocality with projective measurements}
%%%%%%%%%%%%%%%%%%%%%%%%%%%%%%%%%%%%%%%%%%%%%%%%%%%%%%%%%%%%%%%%%%%
\author{Ya Xiao}\email{xiaoya@ouc.edu.cn}
\affiliation{College of Physics and Optoelectronic Engineering, Ocean University of China, Qingdao 266100, People's Republic of China}
\author{Yan-Xin Rong}
\affiliation{College of Physics and Optoelectronic Engineering, Ocean University of China, Qingdao 266100, People's Republic of China}
\author{Xin-Hong Han}
\affiliation{College of Physics and Optoelectronic Engineering, Ocean University of China, Qingdao 266100, People's Republic of China}
\author{Shuo Wang}
\affiliation{College of Physics and Optoelectronic Engineering, Ocean University of China, Qingdao 266100, People's Republic of China}
\author{Xuan Fan}
\affiliation{College of Physics and Optoelectronic Engineering, Ocean University of China, Qingdao 266100, People's Republic of China}
\author{Wei-Chen Li}
\affiliation{College of Physics and Optoelectronic Engineering, Ocean University of China, Qingdao 266100, People's Republic of China}
\author{Yong-Jian Gu}\email{yjgu@ouc.edu.cn}
\affiliation{College of Physics and Optoelectronic Engineering, Ocean University of China, Qingdao 266100, People's Republic of China}

\date{\today}

%%%%%%%%%%%%%%%%%%%%%%%%%%%%%%%%%%%%%%%%%%%%%%%%%%%%%%%%%%%%%%%%%
\begin{abstract}
As a way of saving quantum resources, recycling of Bell nonlocality has been experimentally studied, but restricted to sequential unsharp measurements. However, it has been theoretically shown recently that projective measurements are sufficient for recycling nonlocality [Phys. Rev. Lett. \textbf{129}, 230402 (2022)].  Here, we go beyond unsharp measurement scenarios and experimentally demonstrate the recycling of nonlocal resources with projective measurements. By verifying the violation of Clauser-Horne-Shimony-Holt (CHSH) inequality, we find that three independent parties can recycle the Bell nonlocality of a two-qubit state, whether it is maximally or partially entangled. Furthermore, in the double violation region, the optimal trade-off for partially entangled states can be 11 standard deviations better than that for maximally entangled states.
Our results experimentally eliminate the common misconception that projective measurements are incompatible with the recycling of quantum correlations. In addition, our nonlocality recycling setup does not require entanglement assistance, which is much more experimentally friendly, thus paving the way for the reuse of other kinds of quantum correlations. 
\end{abstract}

\maketitle
 
%%%%%%%%%%%%%%%%%%%%%%%%%%  body  %%%%%%%%%%%%%%%%%%%%%%%%%%%%%%%%%%%%%%%
\section*{I. Introduction}
%%%%%%%%%%%%%%%%%%%%%%%%%%%%%%%%%%%%%%%%%%%%%%%%%%%%%%%%%%%%%%%%%%%%%%%%%
 
Bell non-locality \cite{Bell1964,Brunner2014}, the phenomenon that the results of local measurements performed on distant parties of a composite system can not be explained by local hidden variable theories, plays an important role in device-independent quantum information processing, such as quantum key distribution \cite{Ekert1991,Barrett2005,Liu2022Toward}, randomness expansion and amplification \cite{Acin2007,Pironio2010,Colbeck2012}, quantum secure direct communication \cite{Qi2021,Sheng2022,Zhou2020,Zhou2020}, and communication complexity reduction \cite{Wang2022Xiao,Buhrman2010,Martinez2018,Ho2022}. It is thus of great importance to study how to efficiently reuse this resource. In 2015, Silva \textit{et al.} demonstrated that Bell nonlocality from an entangled pair can be utilized for multiple parties with sequential unsharp measurement of intermediate strength \cite{Silva2015}. Since then, the unsharp measurement method has been widely applied in recycling quantum correlations, such as standard  Bell nonlocality  \cite{Hu2018,Foletto2020,Feng2020,Zhang2021,Ren2022}, network nonlocality   \cite{Mahato2022,Hou2022,Mao2022,Zhang2022,Wang2022,Halder2022}, quantum steering \cite{Sasmal2018,Choi2020,Gupta2021,Han2021,Han2022,Zhu2022,Liu2022}, quantum entanglement \cite{Bera2018,Maity2020,Srivastava2021}, quantum coherence \cite{Datta2018,Hu2022}, and quantum contextuality \cite{Kumari2019,Anwer2021}. Further to this, the maximum number of independent parties that share the quantum nonlocality of a two-qubit entangled state has also been extensively explored   \cite{Shenoy2019,Das2019,Brown2020,Cheng2021,Pandit2022,Srivastava2022}. Moreover, the shared quantum correlations have been used in quantum random access code \cite{Xiao2021,Das2022}, 
randomness certification \cite{Bowles12020,Gupta2022}, and so on. 
 
Usually, to recycle quantum correlations between multiple pairs of parties simultaneously, each sequential party except the last one should perform unsharp measurements. By properly modulating the measurement sharpness, they extract enough information from the system, and reserve enough information for the last party at the same time, thus providing the possibility of reusing the correlation among multiple sequential parties. If projective (unsharp) measurements are performed, the state of the system collapses to one of the eigenstates of the measured observable  \cite{Nielsen2000}, which means that any subsequent measurement on the same system will not provide any additional information about its original state. It seems that the use of projective measurement can not provide aforementioned quantum correlation sharing. However, this is not always true.

Recently, Steffinlongo \textit{et al.} developed a protocol for sharing Bell nonlocality among Alice and a sequence of Bobs when Alice and each Bob stochastically combine three different types of projective measurement strategies \cite{Steffinlongo2022}. The first one is ``basis projection"; % which corresponds to the maximally sharp measurement; 
the second one is ``identity measurement"; and the third one is a ``mixed" strategy with some measurements being basis projections while others identity measurements. Bell non-locality is witnessed by the violation of the CHSH inequality \cite{Clauser1969}. Contrary to the first measurement strategy, the second and the third strategies both prevent a first CHSH inequality violation and enable a next violation. Thus, it is possible to stochastically combine the three strategies to obtain sequential CHSH inequality violations.
 
Up to now, all the experimental researches on Bell nonlocality recycling are restricted to sequential unsharp measurements \cite{Hu2018,Foletto2020,Feng2020}. Here, we report an observation of double CHSH inequality violations for a single pair of entangled polarization photons by randomly combining no more than two different projective measurement strategies. For the maximally entangled state, we find that when stochastically combining basis projection and identity measurement strategies, the correlations shared between Alice and Bob, and the sequential party Charlie both violate the CHSH inequalities. Both violations become even stronger when the combined strategy with identity measurements is replaced by that with mixed measurements. We further verify the optimal trade-off between these two sequential violations, and compare the trade-off with that obtained from partially entangled states. The result shows that, contrasting to the sequential unsharp measurement scenarios, the violations of CHSH inequalities are stronger with partially entangled states compared with that of maximally entangled states. The maximal violation deviation can reach as high as 11 standard deviations.  What is more, partially entangled states with deterministic identity measurement strategy also behave better than the maximally entangled state with optimal stochastically combined measurement strategy. Our results may promote a deeper understanding of the relationship among quantum correlation, quantum measurement and quantum resource recycling.

%%%%%%%%%%%%%%%%%%%%%%%%%%%%%%%%%%%%%%%%%%%%%%%%%%%%%%%%%%%%%%%%%%%%%%%%

\section*{II. Scenario and theory}

%%%%%%%%%%%%%%%%%%%%%%%%%%%%%%%%%%%%%%%%%%%%%%%%%%%%%%%%%%%%%%%%%%%%%%%%
%Fig.1
%%%%%%%%%%%%%%%%%%%%%%%%%%%%%%%%%%%%%%%%%%%%%%%%%%%%%%%%%%%%%%%%%%%%%%%%
\begin{figure}[!htp]
\centering\includegraphics[width=8 cm]{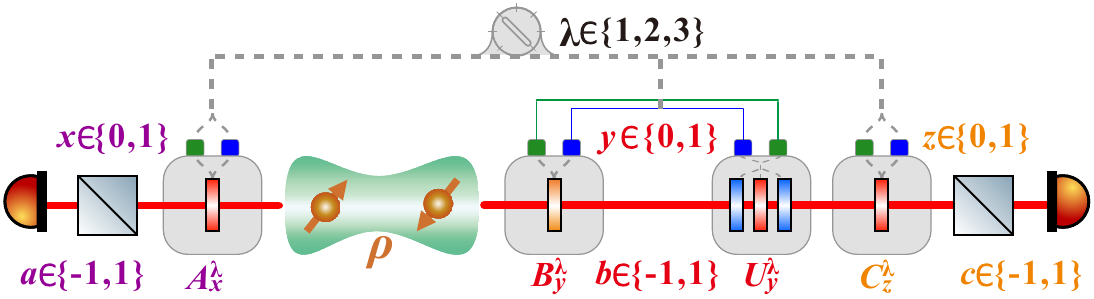}
\caption{Three-party sequential CHSH scenario. A two-qubit entangled state $\rho$ is initially shared between Alice and Bob. After Bob has finished his randomly selected measurement and recorded the outcome, he passes the  post-measurement state to Charlie. All parties are restricted to perform projective measurements and allowed to share strings of classically correlated data $\lambda$, which are used to determine their measurements and unitary operations.} 
\label{model}
\end{figure}
%%%%%%%%%%%%%%%%%%%%%%%%%%%%%%%%%%%%%%%%%%%%%%%%%%%%%%%%%%%%%%%%%%%%%%%%%
%{\it Sequential projective measurement scenario.}---
We focus on a scenario involving three parties, where Alice attempts to share nonlocal correlation with Bob and Charlie using only a two-qubit entangled state $\rho$, one half of which is possessed by Alice, the other half is sequentially sent to Bob and Charlie (see Fig. \ref{model} (a)). In the recycling scenario, each party is restricted to performing two different projective measurement settings, and the Bell nonlocality is verified by the violation of the CHSH inequality \cite{Clauser1969}. There are only three different cases of measurement strategies: both measurement settings are basis projections ($\lambda=1$); both are identity measurements ($\lambda=2$); one is basis projection and the other is identity measurement ($\lambda=3$) \cite{Steffinlongo2022}. Suppose all parties are allowed to share correlated strings of classical data, $\lambda\in\lbrace 1,2,3\rbrace$, and they are used to determine their measurement settings and unitary operations. To begin with, the two-qubit state $\rho$ is shared between Alice and Bob. They perform corresponding dichotomic measurements $A_{x}^{\lambda}$ and $ B_{y}^{\lambda}$ according to the received binary inputs $ x\in\lbrace 0, 1\rbrace$ and  $y\in\lbrace 0, 1\rbrace$, which produce binary outputs $a\in\lbrace -1, 1\rbrace$ and $ b\in\lbrace -1, 1\rbrace $, respectively. Bob then applies a unitary operation $ U_{y}^{\lambda}$ on his post-measurement qubit and relays it to Charlie, who similarly receives a binary input $z\in\lbrace 0, 1\rbrace$, and performs an associated measurement $C_{z}^{\lambda}$, yielding an output $ c\in\lbrace -1, 1\rbrace$. The inputs are statistically uniformly distributed and the sequential parties act independently. Repeating the process several times, the CHSH parameters between Alice and Bob ($S_{AB}^{\lambda}$) as well as between Alice and Charlie ($S_{AC}^{\lambda}$) in a determined measurement case $\lambda$ can be expressed as

%%%%%%%%%%%%%%%%%%%%%%%%%%%%%%%%%%%%%%%%%%%%%%%%%%%%%%%%%%%%%%%%%%%%%%%%%
\begin{center}
\begin{equation}\label{CHSH}
\begin{split}
S_{AB}^{\lambda}&=\sum_{x,y} (-1)^{xy} {\rm Tr}( A_{x}^{\lambda}B_{y}^{\lambda}\rho),\\	
S_{AC}^{\lambda}&=\sum_{x,z}  (-1)^{xz} {\rm Tr}( A_{x}^{\lambda}C_{z}^{\lambda}\rho_{AC}),\\	
 \end{split}
\end{equation}
\end{center}
%%%%%%%%%%%%%%%%%%%%%%%%%%%%%%%%%%%%%%%%%%%%%%%%%%%%%%%%%%%%%%%%%%%%%%%%%
where $\rho_{AC}=\sum_{y,b} (I \otimes U_{y}^{\lambda}\sqrt{B_{b\vert y}^{\lambda} }) \rho (I \otimes U_{y}^{\lambda}\sqrt{B_{b\vert y}^{\lambda} })^{\dagger}/2 $. The violation of CHSH inequality $ S_{AB}^{\lambda} \leq2 $ ($ S_{AC}^{\lambda} \leq2 $) proves the existence of Bell nonlocality between Alice and Bob (Charlie). Consider a shared entangled state $\rho=\vert \psi_{\varphi} \rangle\langle \psi_{\varphi}\vert$ with $ \vert \psi_{\varphi} \rangle={\rm cos} \varphi \vert 00 \rangle +{\rm sin} \varphi \vert 11 \rangle$ and $\varphi\in[ 0, 45^{\circ} ] $, the optimal measurement settings for the aforementioned three cases of measurement strategies are as follows \cite{Steffinlongo2022}.

\textbf{Case 1: basis projection} ($\lambda=1$). Alice measures $A_{0}^{1}=\sigma_{1}$ and  $A_{1}^{1}=\sigma_{3}$. Bob measures $B_{0}^{1}={\rm cos} \phi\sigma_{1}+{\rm sin} \phi \sigma_{3}$ and $B_{1}^{1}={\rm cos} \phi\sigma_{1}-{\rm sin} \phi \sigma_{3}$, and then applies the corresponding unitary operations $U_{0}^{1}=I$ and $U_{1}^{1}=e^{i(\phi-\pi/2)\sigma_{2}}$. Charlie measures $C_{0}^{1}=-C_{1}^{1}={\rm cos} \phi\sigma_{1}+{\rm sin} \phi \sigma_{3}$. $\lbrace \sigma_{1}, \sigma_{2},\sigma_{3} \rbrace $ are Pauli matrices and $ I $ is the identity matrix. This gives $S_{AB}^{1}= 2{\rm cos}\phi{\rm sin}(2\varphi)+2{\rm sin}\phi$ and $S_{AC}^{1}= 2{\rm sin}\phi$. 
 
\textbf{Case 2: identity measurement} ($\lambda=2$). Alice measures $A_{0}^{2}=\sigma_{3}$ and  $A_{1}^{2}=\sigma_{1}$. Bob measures $B_{0}^{2}= B_{1}^{2}=I$, and then applies the corresponding  unitary operations $U_{0}^{2}=U_{1}^{2}=I$. Charlie measures $C_{0}^{2}={\rm cos} \chi\sigma_{1}+{\rm sin} \chi\sigma_{3}$ and  $C_{1}^{2}=-{\rm cos} \chi\sigma_{1}+{\rm sin} \chi\sigma_{3}$, where $\chi={\rm arctan}({\rm csc} (2\varphi)) $. This gives $S_{AB}^2= 2{\rm cos}(2\varphi)$, $S_{AC}^2= 2\sqrt{1+{\rm sin}^2(2\varphi)} $.

\textbf{Case 3: mixed strategy} ($ \lambda=3$).  Alice measures $A_{0}^{3}={\rm cos} \theta\sigma_{1}+{\rm sin} \theta \sigma_{3}$ and $A_{1}^{3}=-{\rm cos} \theta\sigma_{1}+{\rm sin} \theta \sigma_{3}$. Bob measures $B_{0}^{3}=I$ and $  B_{1}^{3}=\sigma_{1} $, and then applies the corresponding unitary operations $U_{0}^{3}=U_{1}^{3}=I$. Charlie measures $C_{0}^{3}= \sigma_{3}$ and  $C_{1}^{3}= \sigma_{1} $. This gives $S_{AB}^3= 2 {\rm sin}(\theta+2\varphi)$, $S_{AC}^3={\rm sin}\theta+ 2{\rm cos}\theta{\rm sin} (2\varphi)$. 

Clearly, case 1 only enables the CHSH inequality violation between Alice and Bob, and both case 2 and case 3 only enable the violation between Alice and Charlie. Thus, we need to use the distribution of the shared randomness $\lbrace p(\lambda)\rbrace_{\lambda=1}^{3}$ to stochastically combine those three cases to obtain the double violations. The CHSH inequalities between Alice and Bob as well as Alice and Charlie can be written as
%%%%%%%%%%%%%%%%%%%%%%%%%%%%%%%%%%%%%%%%%%%%%%%%%%%%%%%%%%%%%%%%%%%%%%%%%
\begin{center}
\begin{equation}\label{CHSH}
\begin{split}
S_{AB}=\sum_{\lambda} p(\lambda) S_{AB}^{\lambda}\leq2,\	
S_{AC}=\sum_{\lambda} p(\lambda) S_{AC}^{\lambda}\leq2.\\	
 \end{split}
\end{equation}
\end{center}
%%%%%%%%%%%%%%%%%%%%%%%%%%%%%%%%%%%%%%%%%%%%%%%%%%%%%%%%%%%%%%%%%%%%%%%%%
The optimal measurement setting $ \lbrace \phi,\chi,\theta  \rbrace $ for the correponding stochastically combined measurement strategy can be obtained by maximizing the minimum value of $ \lbrace S_{AB},S_{AC} \rbrace $. Interesting, we find that, for a suitable choice of $\varphi$, partially entangled states can achieve larger sequential violations than the maximally entangled states. This analysis is detailed in Supplemental Material \cite{SM}.

\section*{III. Experimental setup and results}

%%%%%%%%%%%%%%%%%%%%%%%%%%%%%%%%%%%%%%%%%%%%%%%%%%%%%%%%%%%%%%%%%%%%%%%%%
%Fig.2
%%%%%%%%%%%%%%%%%%%%%%%%%%%%%%%%%%%%%%%%%%%%%%%%%%%%%%%%%%%%%%%%%%%%%%%%%
\begin{figure}[!htp]
\centering
\includegraphics[width=8.5 cm]{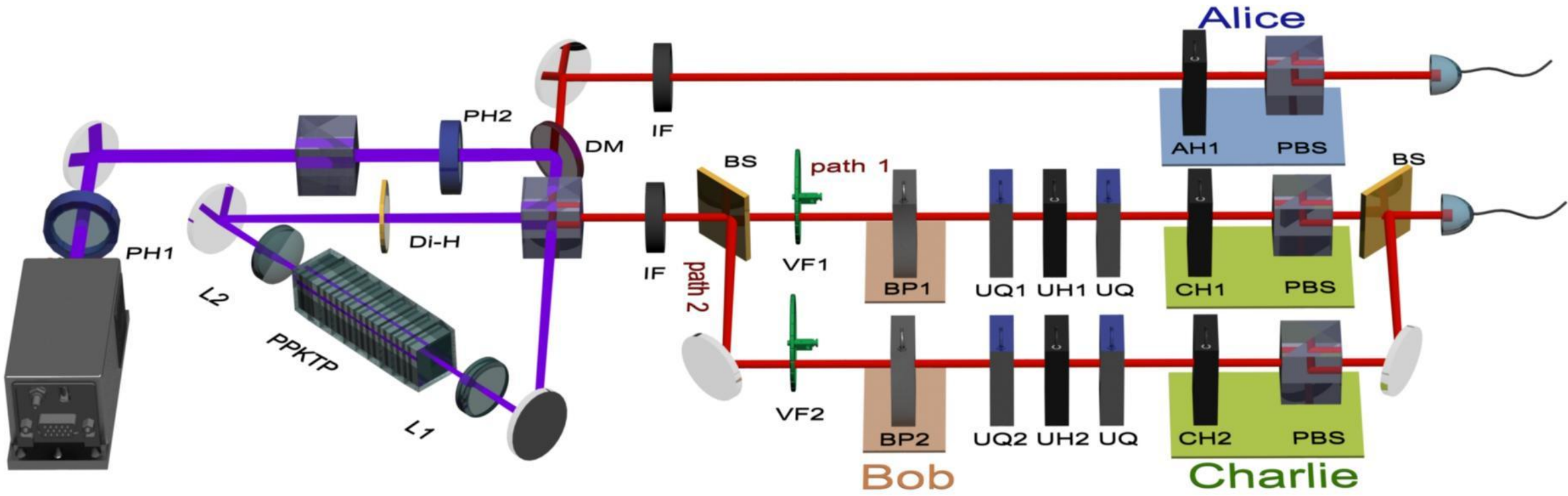}
\caption{Experimental setup. A bright source of polarization entangled photon pairs in a state $\vert \psi_{\varphi} \rangle={\rm cos} \varphi \vert HH \rangle +{\rm sin} \varphi \vert VV \rangle$ ($\varphi \in[0^{\circ}, 45^{\circ}]$) is generated via the spontaneous parametric down-conversion process by pumping a type-II cut PPKTP crystal located in a Sagnac interferometer with an ultraviolet laser at 405 nm. These two photons are filtered by interference filters. One of the photons is sent to Alice, and the other one is subsequently sent to Bob and Charlie by path 1 and path 2. These paths correspond to different measurement strategies and their relative probabilities can be flexibly changed by rotating the variable ND filters. Since Alice, Bob and Charlie only need to measure linear polarizations, hence their setup can be simplified to a polarizer or a composition of a half-wave plate and a polarization beam splitter. The arbitrary qubit unitary operation of Bob can be realized by a composition of two quarter-wave plates, and a half-wave plate. The photons are detected by single photon detectors and the signals are sent for coincidence.
}
\label{setup}
\end{figure}

%{\it Experimental setup and results.}--- 
We focus on investigating the Bell nonlocality recycling scenario where no more than two different projective measurement strategies are combined. Fig. \ref{setup} shows our experimental setup.  A 4 mW pump laser centered at 405 nm is used to pump a 15 mm long  periodically poled potassium titanyl phosphate (PPKTP) crystal in clockwise and counter-clockwise directions to generate a two-qubit polarization-entangled photon state $\vert \psi_{\varphi} \rangle={\rm cos} \varphi \vert HH \rangle +{\rm sin} \varphi \vert VV \rangle$, where $\varphi \in[0^{\circ}, 45^{\circ}]$ \cite{Fedrizzi2007}. $H$ and $V$ represent the horizontal and vertical polarizations, respectively. The parameter $\varphi$ is flexibly changed by the half-wave plate (PH2). The pump light is reflected by a dichroic mirror (DM). Two interference filters (IFs) are used to filter the down-conversion photons. One of the two photons is directly sent to Alice, who uses a half-wave plate (AH1) and a polarization beam splitter (PBS) to perform projective polarization measurement. The other photon is sent to an unbalanced interferometer, and then subsequently sent to Bob and Charlie. In each path, Bob and Charlie carry out a particular measurement strategy. Bob uses linear polarizers (BPs) and Charlie employs half-wave plates (CHs) and polarization beam splitters (PBS) to perform the associated projective measurement. The unitary operation setup, comprised of a quarter-wave plate (UQ), a half-wave plate (UH), and a quarter-wave plate (UQ) on both paths, allows Bob to implement arbitrary single qubit unitary transformation. The relative combining probability between these two paths can be easily controlled by rotating the variable neutral density (ND) filters (VFs).

%%%%%%%%%%%%%%%%%%%%%%%%%%%%%%%%%%%%%%%%%%%%%%%%%%%%%%%%%%%%%%%%%%%%%%%%%
%Fig.3
%%%%%%%%%%%%%%%%%%%%%%%%%%%%%%%%%%%%%%%%%%%%%%%%%%%%%%%%%%%%%%%%%%%%%%%%%
\begin{figure}[!htp]
\centering\includegraphics[width=8 cm]{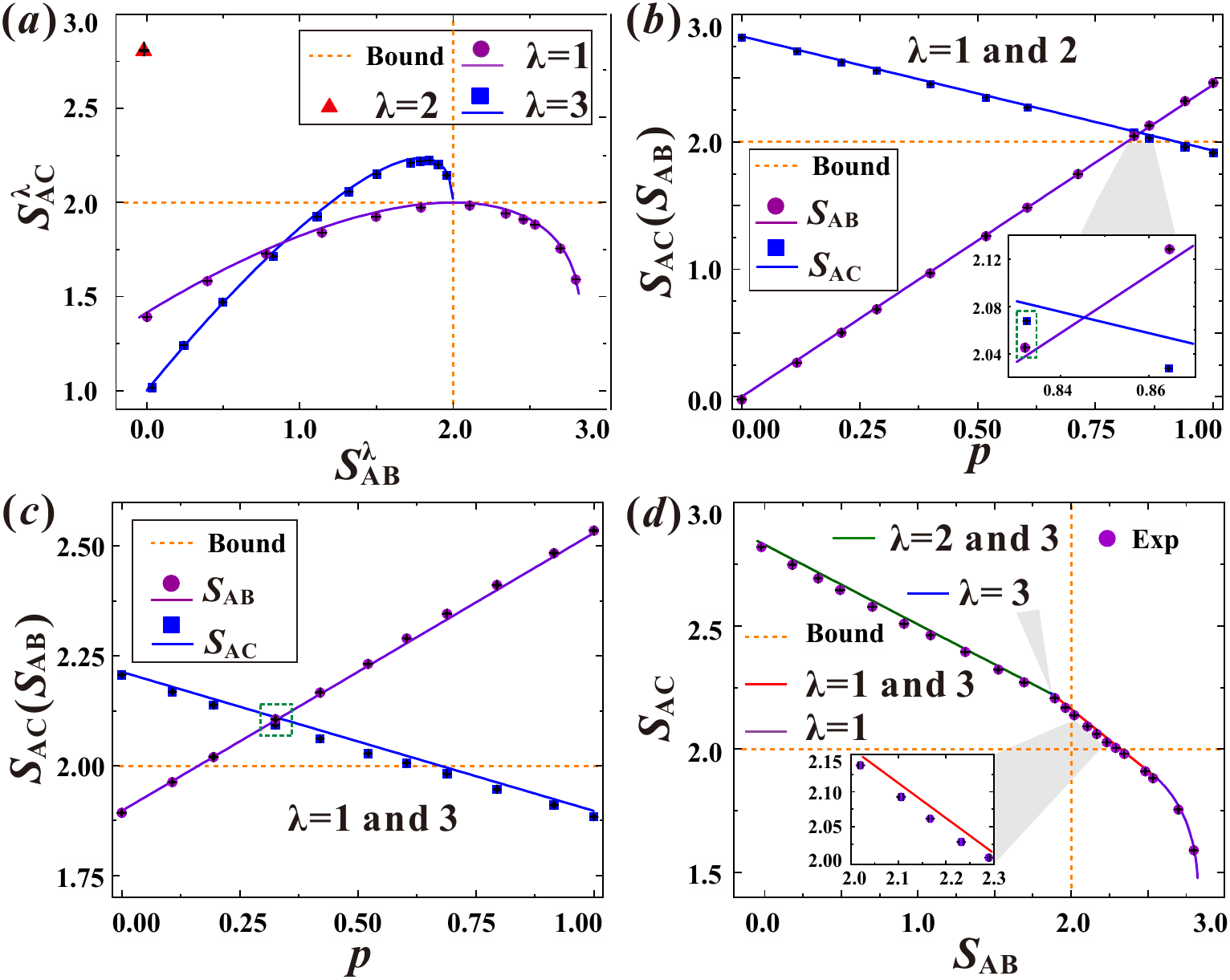}
\caption{The Bell nonlocality recyclability for the maximally entangled state. (a). The optimal trade-off between $S_{AC}^{\lambda}$ and $S_{AB}^{\lambda}$ for case 1 ($\lambda=1$, purple circles), case 2 ($\lambda=2$, red triangle), and case 3 ($\lambda=3$, blue squares) with various measurement settings $\lbrace \phi, \chi, \theta\rbrace $. Only one point is obtained for the case 2 with identity measurement. (b). The CHSH parameters $S_{AB}$ (purple circles) and $S_{AC}$ (blue squares) as a function of probability $p$ when the measurement strategy is a combination of case 1 and case 3. (c). The CHSH parameters $S_{AB}$ (purple circles) and $S_{AC}$ (blue squares) as a function of probability $p$ when the measurement strategy is a combination of cases 1 and 2. Theoretical predictions in (a)-(c) are represented as solid curves with the corresponding colors. (d). The optimal trade-off between $S_{AC}$ and $S_{AB}$. Theoretical predictions and experimental results are represented as curves and purple circles, respectively. The dashed orange lines are the classical bounds of the CHSH inequality. The insets in (b) and (d) show an enlarged view of the double violation results. Error bars are due to the Poissonian counting statistics.  }
\label{maxent}
\end{figure}

We first investigate the Bell nonlocality recycling for a maximally entangled state, i.e., $\varphi=45^{\circ} $. With the optimal measurement setting of each deterministic projective strategy parameterized by $\lambda$, we measured $S_{AC}^{\lambda}$ and $S_{AB}^{\lambda}$ for thirteen values of $\phi$ when $\lambda=1$ and for thirteen values of $\theta$ when $\lambda=3$. And the experimental results of $\lbrace S_{AB}^{\lambda}, S_{AC}^{\lambda}\rbrace$ are obtained at $\chi=45^{\circ} $ when $\lambda=2$. The corresponding trade-offs between $S_{AC}^{\lambda}$ and $S_{AB}^{\lambda}$ are shown in Fig. \ref{maxent}(a). The purple circles, red triangle and blue squares correspond to case 1 ($\lambda=1$), case 2 ($\lambda=2$) and case 3 ($\lambda=3$), respectively. Theoretical predictions are represented as solid curves, and compared with the experimental data.
See Ref. \cite{Steffinlongo2022} and Supplemental Material \cite{SM} for more details. In case 1 and case 3, with the increase of $S_{AB}^{\lambda}$, $S_{AC}^{\lambda}$ first increases and then decreases. And case 1 enables Bell nonlocality shared between Alice and Bob but not between Alice and Charlie, while the opposite is true for case 2 and case 3. To obtain the optimal double violations, we stochastically combine case 1 and case 2 with the optimal measurement setting $\phi=75^{\circ}$ and $\chi=45^{\circ}$. The operations in path 1 and path 2 correspond to the measurements in case 1 and case 2, respectively. By rotating VF1 and VF2, we can further change the probability $p$ of case 1. The CHSH parameters $S_{AB}$ (purple circles) and $S_{AC}$ (blue squares) as a function of $p$ are shown in Fig. \ref{maxent}(b). Clearly, as $p$ increases, $S_{AB}$ increases and $S_{AC}$ decreases. The double violations of CHSH inequality are achieved when $p$ is in the range of $[2/\sqrt{6},(4-2\sqrt{2})/(3-\sqrt{3})]$. Theoretically, $S_{AB}$ and $S_{AC}$ reach their maximal value $ 2\sqrt{2}(\sqrt{3}-1) $ when $ p=(6-2\sqrt{3})/3 $. Experimentally, we get $S_{AB}=2.0451\pm 0.0015$ and $S_{AC}=2.0676\pm 0.0014$ when $p=0.83207\pm 0.0015$, which are both more than 30 standard deviations above the classical bound.  %The results are shown in the dashed green box.  
In addition, we study the Bell nonlocality recycling by stochastically combining case 1 and case 3. The optimal measurements are $\phi=71.57^{\circ}$ and $\theta=18.43^{\circ}$. Fig. \ref{maxent}(c) presents $S_{AB}$ (purple) and $S_{AC}$ (blue) as a function of the probability $ p $ of case 1. The maximum values of the two CHSH parameters $S_{AB}=2.1056\pm 0.0013$ and $S_{AC}=2.0927\pm 0.0013$  are obtained at $p=0.3256\pm 0.0012$, which are both more than 71 standard deviations above the classical bound.  It is clearly that double violations of the CHSH inequality are achieved, and both violations are stronger than that combining case 1 and case 2.  What is more, the double violations enabled range of $ p $ is extended, exhibiting better performance. We further combine the stochastically combined strategy of case 1 and case 3 with other three types of measurement strategies, i.e., a combination of cases 2 and 3 (green curve), deterministic case 3 (blue curve), a combination of cases 1 and 3 (red curve) and deterministic case 1 (purple curve), to obtain the boundary of the attainable CHSH parameters $S_{AB}$ and $S_{AC}$.  This analysis is detailed in Supplemental Material \cite{SM}. The optimal trade-off between $S_{AC}$ and $S_{AB}$ over all range of $S_{AB}$  is shown in Fig. \ref{maxent}(d). Experimental results are in a good agreement with theoretical expectations.

%%%%%%%%%%%%%%%%%%%%%%%%%%%%%%%%%%%%%%%%%%%%%%%%%%%%%%%%%%%%%%%%%%%%%%%%%
%Fig.4
%%%%%%%%%%%%%%%%%%%%%%%%%%%%%%%%%%%%%%%%%%%%%%%%%%%%%%%%%%%%%%%%%%%%%%%%%
\begin{figure}[!htp]
\centering\includegraphics[width=8 cm]{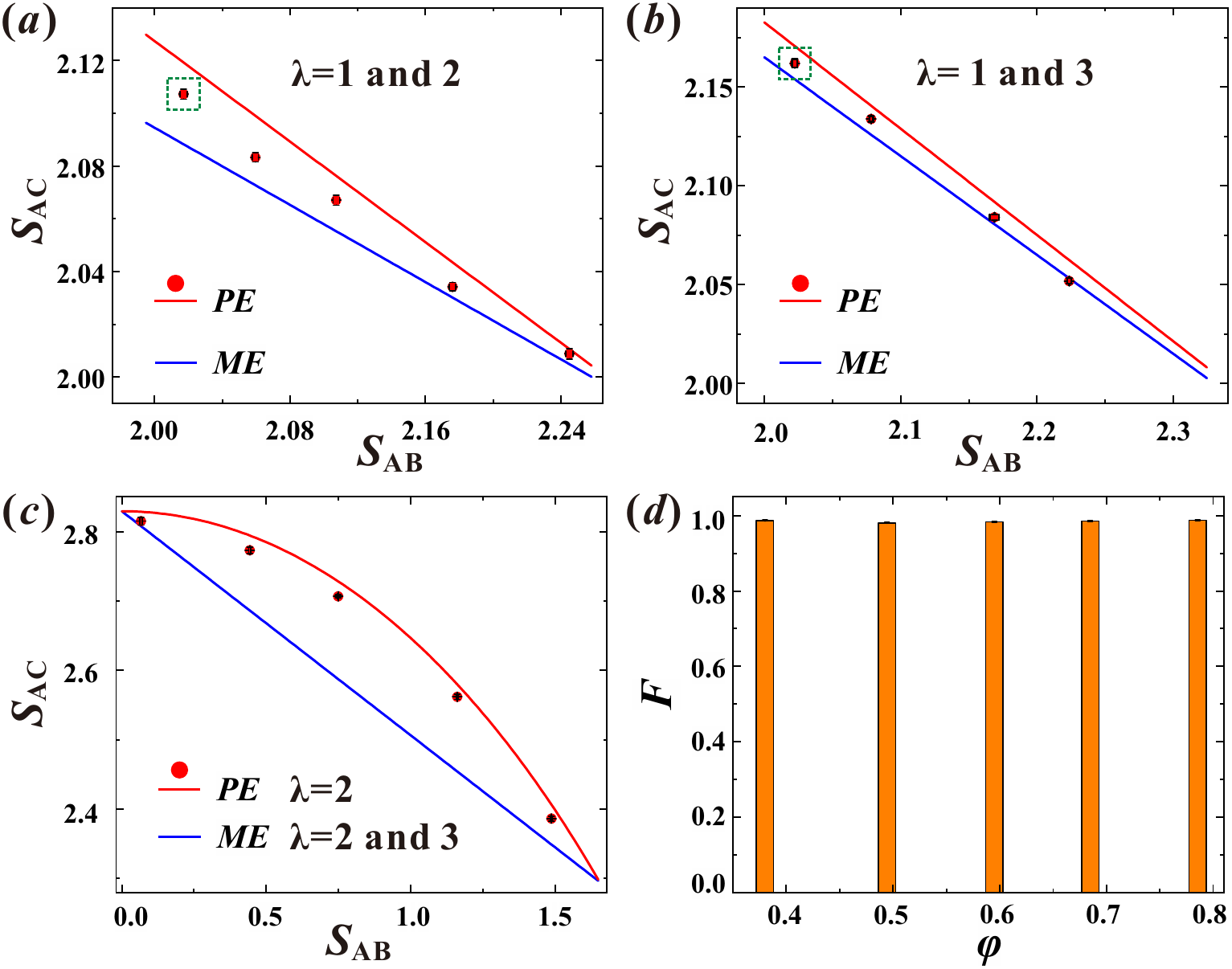}
\caption{The optimal trade-offs between $S_{AC}$ and $S_{AB}$ for various partially entangled states when the measurement strategy is: (a). a combination of case 1 ($\lambda=1$) and case 2 ($\lambda=2$); (b). a combination of case 1 and case 3 ($\lambda=3$); (c). a determined case 2. Experimental results are shown as red circles. Theoretical predictions for partially ($PE$) and maximally ($ME$) entangled states are represented as red and blue curves, respectively. (d). The fidelity of the experimental states shown in (c). Error bars are due to the Poissonian counting statistics.}
\label{compare}
\end{figure}

We further study the sequential Bell nonlocality recycling with partially entangled states.  By varing the probability $ p $ of case 1, we also observe the double violations of CHSH inequality whether the measurement strategy is a stochastic combination of case 1 and case 2 or a combination of case 1 with case 3. The corresponding CHSH parameters $S_{AB}$ and $S_{AC}$ as a function of $p$ are shown in the Supplemental Material, i.e., FIG. S2. And the optimal trade-offs between $S_{AC}$ and $S_{AB}$ satisfying the double violations of CHSH inequality for the states $\varphi=34.08^{\circ}$ and $\varphi=41.48^{\circ}$ are presented in  Fig. \ref{compare}(a) and \ref{compare}(b), respectively. Red solid lines and red circles correspond to theoretical predictions and experimental results. These results are compared with those for the maximally entangled state, whose theoretical results are denoted as the blue solid lines in Fig. \ref{compare}(a) and \ref{compare}(b). We find that, whether stochastically combining case 1 with case 2 or combining case 1 with case 3, for a suitable choice of $\varphi$, the partially entangled states can achieve stronger double violations.  The corresponding $S_{AC}$ shown in the dashed green boxes are about 11 and 5 standard deviations higher than what are possibly achieved based on the maximally entangled state in the above mentioned two cases, respectively. Moreover, by simply applying the deterministic measurement strategy, partially entangled state can also outperform the maximally entangled state. The results for states $\varphi= \lbrace 45^{\circ}, 39.23^{\circ}, 34.08^{\circ}, 28.32^{\circ},21.77^{\circ}\rbrace$ with deterministic measurement case 2 are presented in Fig. \ref{compare}(c). All experimental results exceed the bound of the maximally entangled state. And the difference of $S_{AC}$ reaches the maximum when $\varphi=34.08^{\circ}$. The corresponding state fidelities are further shown in Fig. \ref{compare}(d),  whoes average value is about $0.9853 \pm 0.0012$.

%%%%%%%%%%%%%%%%%%%%%%%%%%%%%%%%%%%%%%%%%%%%%%%%%%%%%%%%%%%%%%%%%%%%%%%%%
\section*{IV. Discussion and conclusion}
%%%%%%%%%%%%%%%%%%%%%%%%%%%%%%%%%%%%%%%%%%%%%%%%%%%%%%%%%%%%%%%%%%%%%%%%%
{\it Discussion and conclusion.}--- In this work, we experimentally demonstrate the Bell nonlocality recycling of a single two-qubit entangled state between Alice, Bob, and Charlie using only projective measurements. In the case of stochastically combining basis projections and identity measurements, we obtain double violations of CHSH inequality for the maximally entangled state. However, in the case of stochastic combinations of basis projections and mixed measurements, both violations become stronger. We further note that some partially entangled states can also achieve stronger double violations. The maximal violation can be 11 standard deviations higher than what is possibly achieved based on the maximally entangled state. What is more, we find partially entangled states with deterministic identity measurement strategy can also behave better than the maximally entangled state with optimal stochastically combined measurement strategy. This is in stark contrast to the standard CHSH scenario and unsharp-measurement-based scheme that partially entangled states are strictly weaker than maximally entangled states. Our results show it is efficient to use only projective measurements to simulate the quantum instruments commonly employed in quantum correlation recycling scenarios, which has never been studied in experiment. In addition, unlike previous unsharp measurement protocols that require entanglement assistance, our recycling scheme only needs qubit projective measurements and unitary transformations, which is more experimentally friendly and can also be employed to investigate other types of quantum correlation sharing.

%%%%%%%%%%%%%%%%%%%%%%%%%%%%%%%%%%%%%%%%%%%%%%%%%%%%%%%%%%%%%%%%%%%%%%%%%
\section*{V. Acknowledgments}
%%%%%%%%%%%%%%%%%%%%%%%%%%%%%%%%%%%%%%%%%%%%%%%%%%%%%%%%%%%%%%%%%%%%%%%%%

The authors thank Armin Tavakoli and Jin-Shi Xu for fruitful discussions and valuable comments. This work was supported by the National Natural Science Foundation Regional Innovation and Development Joint Fund (Grant No. U19A2075), the National Natural Science Foundation of China (Grant No. 12004358), the Fundamental Research Funds for the Central Universities (Grants No. 202041012 and No. 841912027), the Natural Science Foundation of Shandong Province of China (Grant No. ZR2021ZD19), and the Young Talents Project at Ocean University of China (Grant No. 861901013107). 

%%%%%%%%%%%%%%%%%%%%%%%%%%%%%%%%%%%%%%%%%%%%%%%%%%%%%%%%%%%%%%%%%%%%%%%%%
\appendix
\section*{Appendix  }

%%%%%%%%%%%%%%%%%%%%%%%%%%%%%%%%%%%%%%%%%%%%%%%%%%%%%%%%%%%%%%%%%%%%%%%%%
\subsection{A. The optimal measurement settings for stochastically combined measurement strategies}
%%%%%%%%%%%%%%%%%%%%%%%%%%%%%%%%%%%%%%%%%%%%%%%%%%%%%%%%%%%%%%%%%%%%%%%%%%%
In this work, we focus on investigating the recyclability of the Bell nonlocality of a two-qubit state $ \vert \psi_{\varphi} \rangle={\rm cos} \varphi \vert 00 \rangle +{\rm sin} \varphi \vert 11 \rangle$, $\varphi\in[0,45^{\circ}]$. In our recycling scenario, the involved parties, Alice, Bob and charlie, employ a deterministic projective measurement strategy or a combination of two different projective measurement strategies. In the main text, we present the corresponding optimal measurement settings for the deterministic basis projection strategy (case 1, $\lambda=1$), for the deterministic indentity measurement strategy (case 2, $\lambda=2$), and for the deterministic mixed measurement strategy (case 3, $\lambda=3$), respectively. A direct calculation gives the trade-offs between CHSH parameters $S_{AB}^{\lambda}$ and $S_{AC}^{\lambda}$, as shown below \cite{Steffinlongo2022}
\begin{center}
\begin{equation}\label{deter}
\begin{split}
S_{AC}^{1}&=\frac{S_{AB}^{1}+{\rm sin}(2\varphi)\sqrt{4+4{\rm sin}^2(2\varphi)-(S_{AB}^{1})^2}}{1+{\rm sin}^2(2\varphi)},\\	
S_{AC}^{2}&=\sqrt{8-(S_{AB}^{2})^2},\\
S_{AC}^{3}&={\rm sin}(2\varphi)\sqrt{1-\frac{(S_{AB}^{3})^2}{4}}(1-2{\rm cos}(2\varphi)\\
&+\frac{S_{AB}^{3}}{2}(2{\rm sin}^2(2\varphi)+{\rm cos}(2\varphi)).
\end{split}
\end{equation}
\end{center}

\begin{figure*}[!htp]
\centering\includegraphics[width=14 cm]{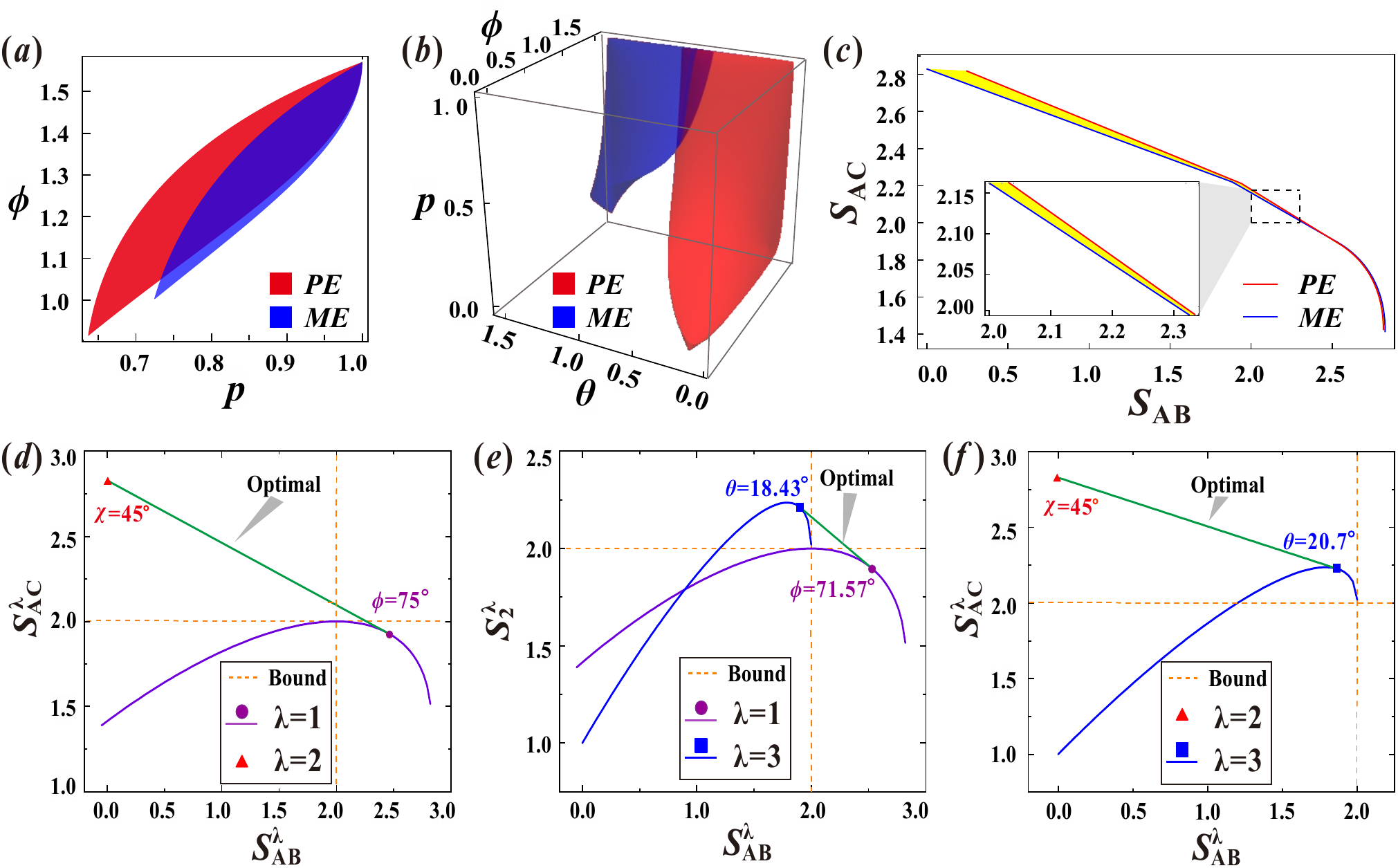}
\caption{(a). The double violation region for the partially entangled state $ \varphi=34.08^{\circ}$ (red, $ PE $) and the maximally entangled state $\varphi=45^{\circ}$ (blue, $ ME $) when the measurement strategy is a combination of case 1 and case 2. (b). The double violation region for the partially entangled state $ \varphi=41.48^{\circ}$ (red, $PE$) and the maximally entangled state $\varphi=45^{\circ}$ (blue, $ME$) when the measurement strategy is a combination of case 1 and case 3. (c). The optimal trade-off between $S_{AB}$ and $S_{AC}$ for the partially entangled state $ \varphi=41.48^{\circ}$ (red curve, $PE$) and the maximally entangled state $ \varphi=45^{\circ}$ (blue curve, $ME$) over all range of $S_{AB}$. The inset shows an enlarged view of the double violation results.
(d). The optimal trade-offs between $S_{AC}^{\lambda}$ and $S_{AB}^{\lambda}$ for the maximally entangled state  $ \varphi=45^{\circ}$ when the measurement strategy involves case 1 ($\lambda=1$) and case 2 ($\lambda=2$). (e). The optimal trade-offs between $S_{AC}^{\lambda}$ and $S_{AB}^{\lambda}$ for the maximally entangled state  $ \varphi=45^{\circ}$ when the measurement strategy involves case 1 ($\lambda=1$) and case 2 ($\lambda=3$). (f). The optimal trade-offs between $S_{AC}^{\lambda}$ and $S_{AB}^{\lambda}$ for the maximally entangled state  $ \varphi=45^{\circ}$ when the measurement strategy involves case 2 ($\lambda=2$) and case 3 ($\lambda=3$). The green lines represent the optimal trade-offs of the corresponding stochastically combined measurement strategies. The purple circles, red triangles and blue squares represent the values of $\lbrace S_{AB}^{\lambda}, S_{AC}^{\lambda}\rbrace$ which are obtained with the optimal measurement settings for case 1, case 2 and case 3, respectively. The dashed orange lines are the classical bounds of the CHSH inequality.}
\label{smtheo}
\end{figure*}

Suppose that the measurement strategy is a stochastic combination of case $s_{i}$ and case $s_{j}$, and the corresponding probabilities are $p(\lambda=s_{i})$ and $p(\lambda=s_{j})$, then CHSH parameters $S_{AB}$ and $S_{AC}$ can be expressed as
%%%%%%%%%%%%%%%%%%%%%%%%%%%%%%%%%%%%%%%%%%%%%%%%%%%%%%%%%%%%%%%%%%%%%%%%%%%
\begin{center}
\begin{equation}\label{CHSH}
\begin{split}
S_{AB}=p(\lambda=s_{i}) S_{AB}^{s_{i}}+p(\lambda=s_{j}) S_{AB}^{s_{j}},\\	
S_{AC}=p(\lambda=s_{i}) S_{AC}^{s_{i}}+p(\lambda=s_{j}) S_{AC}^{s_{j}},\\	
 \end{split}
\end{equation}
\end{center}
%%%%%%%%%%%%%%%%%%%%%%%%%%%%%%%%%%%%%%%%%%%%%%%%%%%%%%%%%%%%%%%%%%%%%%%%%%% 
where $ s_{i}\in\lbrace 1,2,3\rbrace$, $ s_{j}\in\lbrace 1,2,3\rbrace$, $s_{i}\neq s_{j}$ and $ p(\lambda=s_{j})=1-p(\lambda=s_{i})$. There are three types of stochastically combined measurement strategies: a combination of case 1 and case 2, a combination of case 1 and case 3, and a combination of case 2 and case 3. For a given measurement strategy, the violation of CHSH inequality $ S_{AB} \leq2 $ ($ S_{AC} \leq2 $) proves the existence of Bell nonlocality between Alice and Bob (Charlie). Interesting, we find that, for a suitable choice of $\varphi$, partially entangled states can achieve larger sequential violations than the maximally entangled states.  Fig. \ref{smtheo}(a) shows the Bell nonlocality regions for states $\varphi=34.08^{\circ}$ and $\varphi=45^{\circ}$  parameterized by the measurement setting parameter $\phi$ and the probability $p(\lambda=1)$ (abbreviated as $p$) when $p(\lambda=2)=1-p$ and $p(\lambda=3)=0$. Fig. \ref{smtheo}(b) shows the Bell nonlocality regions for states $\varphi=41.48^{\circ}$ and $\varphi=45^{\circ}$ parameterized by the measurement setting parameter $\phi$ and the probability $p$ when $p(\lambda=2)=0$ and $p(\lambda=3)=1-p$. The red and blue regions in (a) and (b) correspond to the double CHSH inequality violation regions for the partially and maximally entangled states, respectively. Obviously, the double violation regions for partially entangled states are larger than that for maximally entangled states.

In the following, we will deduce the optimal measurement settings for aforementioned three stochastically combined measurement strategies by analyzing the optimal trade-offs between $S_{AB}$ and $S_{AC}$. Notice that in all three deterministic cases, the trade-offs between $S_{AC}^{\lambda}$ and $S_{AB}^{\lambda}$ are concave. In the case of shared randomness, the optimal trade-off between $S_{AC}$ and $S_{AB}$ can be expressed as $S_{AC}=k\cdot S_{AB}+S_{0}$, where $ k=\dfrac{S_{AC}^{s_{2}}-S_{AC}^{s_{1}}}{S_{AB}^{s_{2}}-S_{AB}^{s_{1}}}$. To keep both CHSH parameters $S_{AB}$ and $S_{AC}$ as large as possible above the optimal classical bound, the best way is to set measurement settings to maximize the minimum value of $\lbrace S_{AC}, S_{AB} \rbrace $. It is easy to find that, from Eq. (\ref{deter}), $S_{AC}^{\lambda}$ decreases as $S_{AB}^{\lambda}$ increases. Similarly, $S_{AC}$ decreases as $S_{AB}$ increases. And the maximum value of the minimum value of $\lbrace S_{AC}, S_{AB} \rbrace $ is obtained at $S_{AC}= S_{AB} $. This is the main idea for us to obtain the optimal measurement settings of the stochastically combined projective strategy. Substituting $ k=\dfrac{S_{AC}^{s_{2}}-S_{AC}^{s_{1}}}{S_{AB}^{s_{2}}-S_{AB}^{s_{1}}}$ and $S_{AC}= S_{AB} $ into $S_{AC}=k\cdot S_{AB}+S_{0}$, we get the objective function  $S_{AC}= S_{AB}= S_{0}/(1-k)$.
 
As mentioned in the main text, $S_{AB}^{1}= 2{\rm cos}\phi{\rm sin}(2\varphi)+2{\rm sin}\phi$, $S_{AC}^{1}= 2{\rm sin}\phi$, $S_{AB}^2= 2{\rm cos}(2\varphi)$, $S_{AC}^2= 2\sqrt{1+{\rm sin}^2(2\varphi)}$, $S_{AB}^3= 2 {\rm sin}(\theta+2\varphi)$ and $S_{AC}^3={\rm sin}\theta+ 2{\rm cos}\theta{\rm sin} (2\varphi)$. For an arbitrary combination strategy  $ \lbrace s_{1},s_{2}\rbrace $, one can obtaine the optimal state and measurement parameters $ \lbrace\varphi, \phi,\chi,\theta\rbrace $ by finding the maximum value of $ S_{0}/(1-k)$, where $\chi={\rm arctan}({\rm csc} (2\varphi)) $. And for a particular state $\varphi $, the optimal measurement parameters $\lbrace \phi,\chi,\theta\rbrace $ can also be obtained by finding the maximum value of $ S_{0}/(1-k)$. The results for a maximally entangled qubit state $\varphi=45^{\circ}$ are shown in Fig. \ref{smtheo}(d)-(f). With the optimal measurement settings, we can obtaine the results of $\lbrace S_{AB}^{\lambda}, S_{AC}^{\lambda}\rbrace$, which are marked as purple circles, red triangles and blue squares for case 1, case 2 and case 3, respectively. Theoretically, $ S_{AB}^{2}=0$, $S_{AC}^{2}=2\sqrt{2}$,  $\lbrace S_{AB}^{1}, S_{AC}^{1}\rbrace$ and $\lbrace S_{AB}^{3}, S_{AC}^{3}\rbrace$ are the tangent points of the corresponding tangents  \cite{Steffinlongo2022}. The optimal trade-offs between $S_{AB}$ and $S_{AC}$ can be obtained by stochastically combining the corresponding optimal measurement settings, which are represented as green lines. Clearly, both $S_{AB}^{\lambda}$ and $S_{AC}^{\lambda}$ are above the classical bound whether the measurement strategy is the combination of case 1 and case 2 or the combination of case 1 and case 3. However, when case 2 and case 3 are stochastically combined, only $S_{AC}^{\lambda}$ can beat the classical bound.

By comparing the trade-offs of the above mentioned three deterministic projective measurement strategies and three stochastically combined projective measurement strategies, we can further obtain the boundary of the attainable CHSH parameters $\lbrace S_{AB}, S_{AC}\rbrace$. The optimal trade-off between $S_{AC}$ and $S_{AB}$ over all range of $S_{AB}$ ($ 0\leq S_{AB}\leq 2\sqrt{2}$) is divided into four parts. The first one is a combination of case 2 and case 3, the second one is a deterministic case 3, the third one is a combination of case 1 and case 3, and the forth one is a deterministic case 1. The results for states $\varphi=41.48^{\circ}$ (red curve) and $\varphi=45^{\circ}$ (blue curve) are shown in Fig. \ref{smtheo} (c).  The inset shows an enlarged view of the double violation results. It is clear that the partially entangled state can outperform the maximally entangled state.

%%%%%%%%%%%%%%%%%%%%%%%%%%%%%%%%%%%%%%%%%%%%%%%%%%%%%%%%%%%%%%%%%%%%%%%%%
%Fig.5
%%%%%%%%%%%%%%%%%%%%%%%%%%%%%%%%%%%%%%%%%%%%%%%%%%%%%%%%%%%%%%%%%%%%%%%%%

%%%%%%%%%%%%%%%%%%%%%%%%%%%%%%%%%%%%%%%%%%%%%%%%%%%%%%%%%%%%%%%%%%%%%%%%%

%%%%%%%%%%%%%%%%%%%%%%%%%%%%%%%%%%%%%%%%%%%%%%%%%%%%%%%%%%%%%%%%%%%%%%%%%%%

%%%%%%%%%%%%%%%%%%%%%%%%%%%%%%%%%%%%%%%%%%%%%%%%%%%%%%%%%%%%%%%%%%%%%%%%%%%
\subsection{B. The detailed method for determining the experimental parameter $p(\lambda)$}  
%%%%%%%%%%%%%%%%%%%%%%%%%%%%%%%%%%%%%%%%%%%%%%%%%%%%%%%%%%%%%%%%%%%

As shown in Fig. 2 in the main text, path 1 and path 2 are used to implement different deterministic projective measurement strategies. Suppose the measurement strategy implemented in path $i$ is $s_{i} $, where $ i\in\lbrace 1,2\rbrace $, $s_{i}\in\lbrace 1,2,3\rbrace $ and $s_{1}\neq s_{2}  $. The corresponding probability $p(\lambda=s_{i})$ is determined as below.

(1). Firstly, we block path 2. A complete measurement basis of $\lbrace HH, HV,VH,VV\rbrace $ is carried out with the counting rate denoted by $N_{1}$, $N_{2}$, $N_{3}$ and $N_{4}$, respectively. We obtain the total photon count is $P_{1}={N_{1}+N_{2}+N_{3}+N_{4}} $.

(2). Secondly, we block path 1.  A complete measurement basis of $\lbrace HH, HV,VH,VV\rbrace $ is carried out with the counting rate denoted by $M_{1}$, $M_{2}$, $M_{3}$ and $M_{4}$, respectively. We obtain the total photon count is $P_{2}={M_{1}+M_{2}+M_{3}+M_{4}}$.

(3). Then, we get $p(\lambda=s_{1})= P_{1}/(P_{1}+P_{2})$ and $p(\lambda=s_{2})=1-p(\lambda=s_{1})= P_{2}/(P_{1}+P_{2})$.

%%%%%%%%%%%%%%%%%%%%%%%%%%%%%%%%%%%%%%%%%%%%%%%%%%%%%%%%%%%%%%%%%%%%%%%%%%%
\subsection{C. More experimental results}
%%%%%%%%%%%%%%%%%%%%%%%%%%%%%%%%%%%%%%%%%%%%%%%%%%%%%%%%%%%%%%%%%%%%%%%%%%%%

\begin{figure*}[!htp]
\centering\includegraphics[width=15cm]{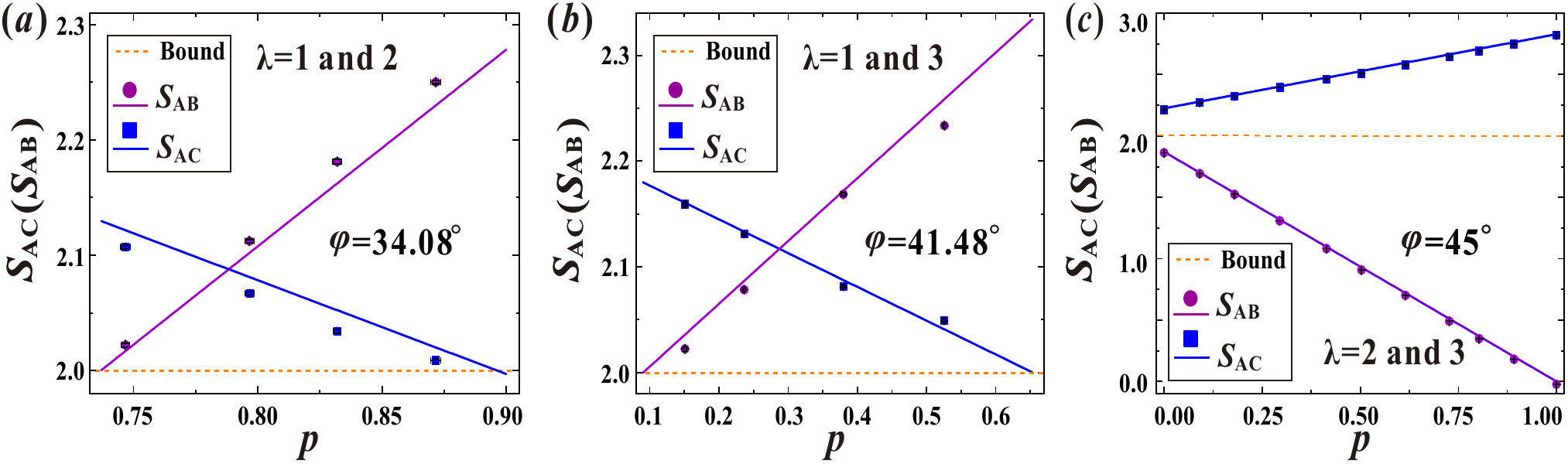}
\caption{ (a). The CHSH parameters $S_{AB}$ (purple circles) and $S_{AC}$ (blue squares) as a function of probability $p(\lambda=1)$ (abbreviated as $p$)  when the initial state is $\varphi=34.08^{\circ} $  and the measurement strategy is a combination of case 1 $ (\lambda=1) $ and case 2 $ (\lambda=2) $. (b).  The CHSH parameters $S_{AB}$ (purple circles) and $S_{AC}$ (blue squares) as a function of probability $p(\lambda=1)$ (abbreviated as $p$) when  the initial state is $\varphi=41.48^{\circ} $ and the measurement strategy is a combination of case 1  and case 3 $ (\lambda=3) $. (c).  The CHSH parameters $S_{AB}$ (purple circles) and $S_{AC}$ (blue squares) as a function of probability $p(\lambda=2)$ (abbreviated as $p$) when  the initial state is $\varphi=45^{\circ} $ and the measurement strategy is a combination of case 2  and case 3. Theoretical predictions in (a)-(c) are represented as solid lines with the corresponding colors. The dashed orange lines are the classical bounds of the CHSH inequality. Error bars are due to the Poissonian counting statistics.}
\label{smexp}
\end{figure*}

For a given stochastically combined measurement strategy, using the optimal measurement settings introduced in section I, we can obtain the relation between the CHSH parameter and the combination probability.  Fig. \ref{smexp} (a) presents the CHSH parameters $S_{AB}$ and $S_{AC}$ as a function of probability $p(\lambda=1)$ (abbreviated as $p$) when the initial state is $\varphi=34.08^{\circ}$ and the measurement strategy is a combination of case 1 $(\lambda=1)$ and case 2 $ (\lambda=2) $. Fig. \ref{smexp} (b) presents the CHSH parameters $S_{AB}$ and $S_{AC}$  as a function of probability $p(\lambda=1)$ (abbreviated as $p$) when the initial state is $\varphi=41.48^{\circ} $  and the measurement strategy is a combination of case 1 and case 3 $(\lambda=3)$. Fig. \ref{smexp} (c) presents the CHSH parameters $S_{AB}$ and $S_{AC}$ as a function of probability $p(\lambda=2)$ (abbreviated as $p$) when the initial state is $\varphi=45^{\circ}$ and the measurement strategy is a combination of case 2 and case 3. Experimentally measured CHSH parameters are shown as symbols, where the purple circles and blue squares correspond to $S_{AB}$ and $S_{AC}$, respectively. Theoretical predictions are also represented as curves, and compared with the experimental data. We find that, whether stochastically combining case 1 with case 2 or combining case 1 with case 3, for a suitable choice of $\varphi$, the partially entangled states can also achieve the double violations of CHSH inequality.  However, when case 2 and case 3 are stochastically combined, only $S_{AC}$ can beat the classical bound even if the initial state is  maximally entangled.

%%%%%%%%%%%%%%%%%%%%%%%%%%%%%%%%%%%%%%%%%%%%%%%%%%%%%%%%%%%%%%%%%%%%%%%%%
%Fig.2
%%%%%%%%%%%%%%%%%%%%%%%%%%%%%%%%%%%%%%%%%%%%%%%%%%%%%%%%%%%%%%%%%%%%%%%%%

%%%%%%%%%%%%%%%%%%%%%%%%%%%%%%%%%%%%%%%%%%%%%%%%%%%%%%%%%%%%%%%%%%%%%%%%%

\end{document}